\documentclass[conference]{IEEEtran}
\IEEEoverridecommandlockouts
\usepackage{cite}
\usepackage{amsmath,amssymb,amsfonts}
\usepackage{algorithm}
\usepackage{algpseudocode}
\usepackage{graphicx}
\usepackage{textcomp}
\usepackage{bm}
\usepackage{array}

\usepackage[utf8]{inputenc}

\usepackage[caption=false,font=footnotesize,labelfont=rm,textfont=rm,subrefformat=parens,labelformat=parens]{subfig}

\usepackage{multirow}
\usepackage{CJK}
\usepackage{cleveref}
\usepackage{booktabs}
\usepackage{steinmetz}
\usepackage{url}

\usepackage[table,xcdraw]{xcolor}

\def\BibTeX{{\rm B\kern-.05em{\sc i\kern-.025em b}\kern-.08em
    T\kern-.1667em\lower.7ex\hbox{E}\kern-.125emX}}
\begin{document}

\title{Movable Antenna-Equipped UAV for Data Collection in Backscatter Sensor Networks: A \\Deep Reinforcement Learning-based Approach}
 
\author{
\IEEEauthorblockN{
Yu Bai\IEEEauthorrefmark{1},
Boxuan Xie\IEEEauthorrefmark{1},
Ruifan Zhu\IEEEauthorrefmark{2},
Zheng Chang\IEEEauthorrefmark{3}\IEEEauthorrefmark{4}, and
Riku J\"antti\IEEEauthorrefmark{1}}
\IEEEauthorblockA{\IEEEauthorrefmark{1}Department of Information and Communications Engineering, Aalto University, Espoo, Finland}
\IEEEauthorblockA{\IEEEauthorrefmark{2}Department of Engineering, King's College London, UK}
\IEEEauthorblockA{\IEEEauthorrefmark{3}School of Computer Science and Engineering, University of Electronic Science and Technology of China, Chengdu, China}
\IEEEauthorblockA{\IEEEauthorrefmark{4}Faculty of Information Technology, University of Jyv\"askyl\"a, Jyv\"askyl\"a, Finland}
}

\maketitle

\begin{abstract}
Backscatter communication (BC) becomes a promising energy-efficient solution for future wireless sensor networks (WSNs). Unmanned aerial vehicles (UAVs) enable flexible data collection from remote backscatter devices (BDs), yet conventional UAVs rely on omni-directional fixed-position antennas (FPAs), limiting channel gain and prolonging data collection time. 
To address this issue, we consider equipping a UAV with a directional movable antenna (MA)
with high directivity and flexibility. 
The MA enhances channel gain by precisely aiming its main lobe at each BD, focusing transmission power for efficient communication. 
Our goal is to minimize the total data collection time by jointly optimizing the UAV’s trajectory and the MA’s orientation. We develop a deep reinforcement learning (DRL)-based strategy using the azimuth angle and distance between the UAV and each BD to simplify the agent’s observation space. 
To ensure stability during training, we adopt Soft Actor-Critic (SAC) algorithm that balances exploration with reward maximization for efficient and reliable learning.
Simulation results demonstrate that our proposed MA-equipped UAV with SAC outperforms both FPA-equipped UAVs and other RL methods, achieving significant reductions in both data collection time and energy consumption.
\end{abstract}


\section{Introduction}
\label{sec:Introduction}

    Large-scale deployment of energy-efficient wireless sensor networks (WSNs) is crucial in the 5G era and beyond. Among Internet of Things (IoT) technologies, backscatter communication (BC) stands out for its energy efficiency, allowing backscatter devices (BDs) to communicate by reflecting existing RF signals with power consumption as low as microwatts-level, bypassing the need for batteries~\cite{duan2020ambient, toro2022survey}. This low-power feature is vital for large-scale WSNs where battery replacement for billions of sensor nodes is impractical. 
    As IoT applications expand, the deployment of BDs presents an energy-efficient alternative for continuous environmental monitoring in remote areas.
    However, a key constraint of BDs is their limited communication range, which presents challenges for data collection in large sensor networks.

    To overcome the short communication range issue, using unmanned aerial vehicles (UAVs) for BD data collection has emerged as a viable solution~\cite{yang2021uav, bletsas2021uav, mao2023uav, riku2024uav}.
    UAV trajectory optimization has been shown to enhance data collection efficiency, yet most existing studies assume that UAVs use omni-directional fixed-position antennas (FPAs) with limited directivity and gain~\cite{zhang2021hierarchical,li2020joint}, resulting in low channel gain and high energy consumption.  
    To improve this, recent studies have proposed movable antenna (MA) systems, which offer a flexible solution for enhancing wireless communications~\cite{zhu2023movable}.
    Superior to FPAs, MAs can dynamically adjust their positions and orientations to adapt to varying environmental conditions, achieving greater spatial diversity and minimizing interference in UAV-assisted data collection scenarios~\cite{tang2024uav}. 
    This adaptability makes MAs a powerful enhancement for UAVs, enabling optimized channel gain in dynamic environments~\cite{zhu2024movable}. Although the MA-equipped UAV offers a more flexible solution for BD data collection, its high mobility introduces greater challenges in optimizing the data collection strategy.

    The optimization challenges in complex, dynamic environments involving multiple items result in non-convex problems and intricate problem formulations. This complexity has driven recent research to leverage deep reinforcement learning (DRL) for its advanced decision-making capabilities within UAV networks~\cite{bai2023towards}. By setting UAV-BD interactions and performance targets, DRL can iteratively enhance system efficiency. Existing DRL studies focus on trajectory optimization of FPA-equipped UAVs \cite{sun2021aoi, zhang2021hierarchical}, leaving open questions on the MA's role in advancing UAV data collection and optimization.
    To fill this gap, in this paper, we investigate a novel scenario where a UAV is equipped with a directional MA to facilitate efficient data collection from BDs. The contributions of this paper are as follows:
    \begin{itemize}
    \item We consider an MA-equipped UAV system that is suitable for BD data collection in backscatter sensor networks. This system leverages both the flexible mobility of the UAV and the reorientation capability of the MA to enhance signal gain and reduce data collection time and energy consumption.
    \item We propose a DRL-based approach to jointly optimize the UAV’s movement and MA reorientation, aiming to minimize data collection time. A soft actor-critic framework is used to enable comprehensive exploration of UAV and MA actions while maintaining stable convergence throughout training.
    \end{itemize}

    The remainder of this paper is organized as follows: Section \ref{sec:System_Model} presents the system model and problem formulation of the data collection system. Section \ref{sec:DRL} details the proposed DRL-based algorithm. Section \ref{sec:Numerical_Results} provides the simulation results. Finally, section \ref{sec:Conclusion} concludes the paper.

\section{System Model}
\label{sec:System_Model}
    \begin{figure}[!t]
    \centering
        \includegraphics[width=0.97\columnwidth]{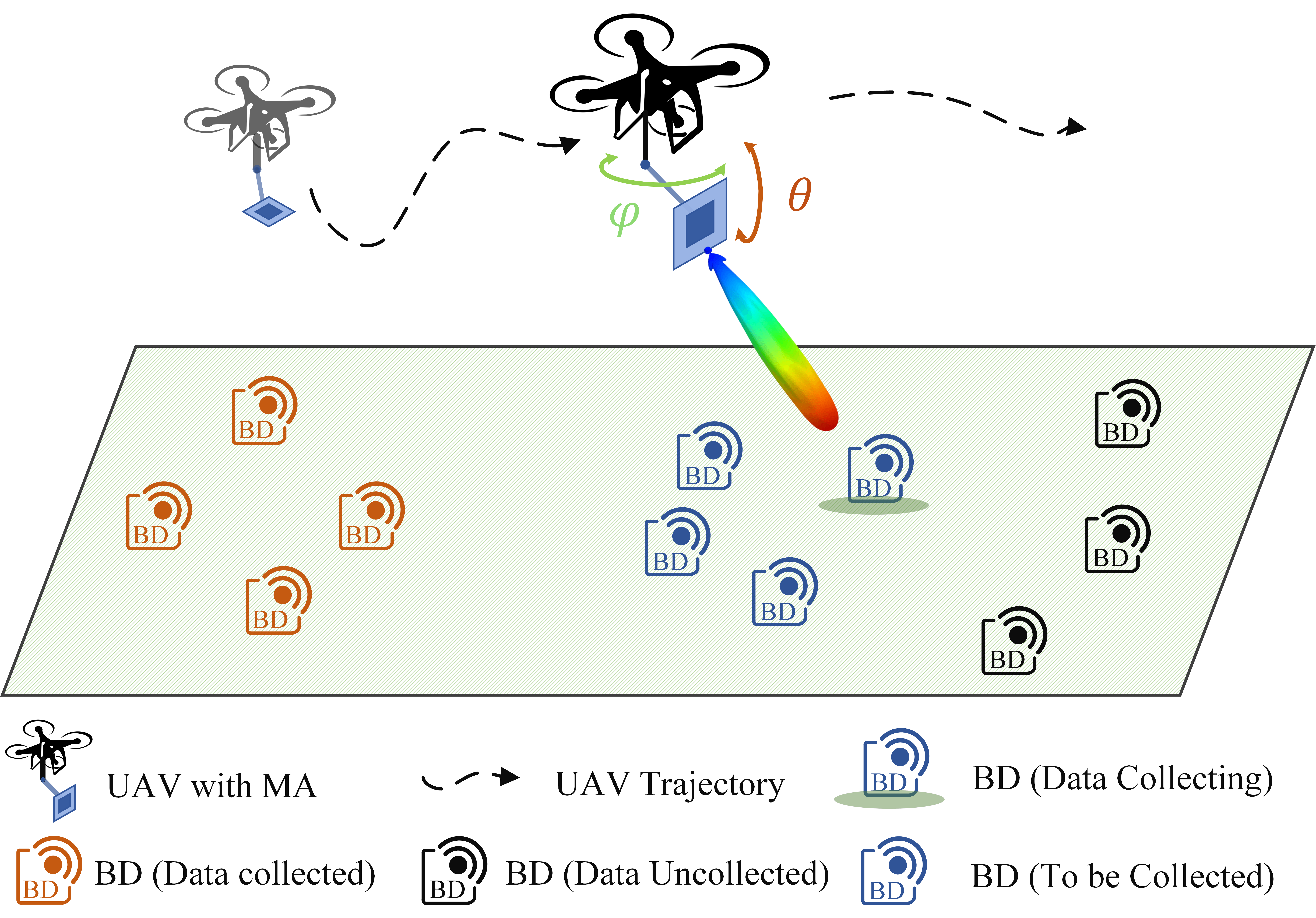}
        \caption{MA-equipped UAV for data collection from BDs.} 
        \label{fig:system_model}
    \vspace{-15pt}
    \end{figure}
    We consider a scenario where a single UAV is deployed to gather data from $K$ BDs within a target area. The target area is modeled as a square with side length $L$. The UAV's position at time $t$ is denoted by $[x(t), y(t), h(t)]$, where $0 \leq x(t), y(t) \leq L$, $h(t) = H$ denotes the fixed altitude of the UAV. The UAV has prior knowledge of the locations of the BDs, which are denoted by the set of coordinates $\bm{p} = \left[ p_{x}^{k}, p_{y}^{k}, 0 \right], k \in \mathcal{K} \triangleq { 1, 2, \dots, K }$. Each BD is self-powered, relying on energy harvested from solar and radio frequency (RF) sources~\cite{toro2022survey}. The UAV provides the RF carrier signal necessary for communication activation. In this system, data transmission is facilitated using \textit{monostatic} backscatter communication~\cite{griffin2009linkbudget}, where the UAV serves both as the RF carrier generator and the receiver of the backscattered signal. 

\subsection{MA-assisted UAV-BD Communication}
    The UAV is equipped with a directional MA. With the directional antenna characteristic, the channel gain in backscatter communication can be significantly improved by adjusting the MA's main lobe orientation $(\theta, \varphi)$, where $\theta \in [0,\pi/2)$ denotes the elevation angle and $\varphi \in [0, 2\pi)$ denotes the azimuth angle, as shown in Fig.~\ref{fig:system_model}. When the UAV collects data from the $k$-th BD, the optimal orientation of the MA to maximize channel gain is determined by setting the elevation and azimuth angles $(\theta_k, \varphi_k)$ as follows
    \begin{equation}
    \label{eqn:elevation angle}
        \theta_k = \arctan \left( \frac{h(t)}{\sqrt{(p_x^k - x(t))^2 + (p_y^k - y(t))^2}} \right),
    \end{equation}
    \begin{figure}
    \vspace{-15pt}
    \end{figure}
    \begin{equation}
    \label{eqn:azimuth angle}
        \varphi_k = \operatorname{atan2}(p_y^k - y(t), p_x^k - x(t)),
    \end{equation}
    \noindent where the $\operatorname{atan2}(y, x)$ function calculates the arctangent of $\frac{y}{x}$ while accounting for the signs of both $x$ and $y$, providing the correct angle across all four quadrants in the range $[0,2\pi]$. This orientation allows the UAV to direct the main lobe of the antenna pattern directly at each BD, maximizing RF power transfer for efficient backscatter communication. Equipped with the MA, the UAV can hover at a position to collect data from multiple BDs. For each BD that meets the received signal strength requirements, the UAV reorients the MA to collect data from all qualified BDs sequentially. The UAV reader transmitted RF carrier signal with a frequency of $f_{\text{c}}$ illuminates the BD that employs ON-OFF keying (OOK) baseband modulation. 
    The baseband-modulated message of each BD consists of preambles with a unique identifier and sensor data. Once data collection is complete for all qualified BDs at the current position, the UAV proceeds to the next location and repeats this process. This cycle continues until data from all BDs in the area have been collected. \par
\subsection{Channel Model for UAV-BD Communications}
    The communication between a single UAV and multiple BDs on the ground can be described using a probabilistic model that accounts for both line-of-sight (LoS) and non-line-of-sight (NLoS) conditions~\cite{al2014optimal}. This model incorporates the probabilities of both LoS and NLoS occurrences to calculate the overall path loss between the UAV and the $k$-th BD.

    \subsubsection{Probability-Based LoS Model}
        The total path loss $\overline{L}(t)$ at time $t$ for the communication link between the UAV and the $k$-th BD is calculated as the weighted sum of the LoS and NLoS path losses, expressed by
        \begin{equation}
        \label{eqn:1}
            \overline{L_{k}}(t) = \rho_{\text{LoS},k}(t) \cdot {{L}}_{\text{LoS},k}(t) + \rho_{\text{NLoS},k}(t) \cdot {{L}}_{\text{NLoS},k}(t),
        \end{equation}
        where $\rho_{\text{LoS},k}(t)$ and $\rho_{\text{NLoS},k}(t)$ represent the probabilities of having a LoS or NLoS link, respectively, while ${{L}}_{\text{LoS},k}(t)$ and ${{L}}_{\text{NLoS},k}(t)$ denote the path loss for LoS and NLoS conditions.
        The path loss for both LoS and NLoS conditions incorporates free-space propagation loss and additional environmental losses, which are defined as
        \begin{equation}
            {L}_{\text{LoS},k}(t) = 20 \log_{10} \left( \frac{4 \pi r_k }{\lambda_{\text{c}} \sqrt{G_{\text{TR}} G_{\text{BD},k}}} \right) + \eta_{\text{LoS}},
        \end{equation}
        \begin{equation}
            {L}_{\text{NLoS},k}(t) = 20 \log_{10} \left( \frac{4 \pi r_k}{\lambda_{\text{c}} \sqrt{G_{\text{TR}} G_{\text{BD},k}}}   \right) + \eta_{\text{NLoS}},
        \end{equation}
        where $\lambda_{\text{c}}$ is the wavelength corresponding to the carrier frequency $f_{\text{c}}$, 
        $G_{\text{TR}}$ and $G_{\text{BD},k}$ are antenna gains of UAV and the $k$-th BD, respectively,
        $r_k$ represents the distance between the UAV and the $k$-th BD with
        \begin{equation}
            r_k(t) = \left[ (x(t) - p_{x}^{k})^2 + (y(t) -  p_{y}^{k})^2 + H^2 \right]^{\frac{1}{2}},
        \end{equation}
        and $\eta_{\text{LoS}}$, $\eta_{\text{NLoS}}$ are the additional losses under LoS and NLoS conditions, respectively.
        The probability of a LoS link between the UAV and the $k$-th BD is modeled as a function of the elevation angle $\theta$, and is given by
        \begin{equation}
            \rho_{\text{LoS}}(t) = \frac{1}{1 + \varrho \exp[-\beta(\frac{\pi}{2}-\theta - \varrho)]},
        \end{equation}
        where $\varrho$ and $\beta$ are environment-dependent constants (e.g., urban, suburban, or rural settings). The NLoS probability is then $\rho_{\text{NLoS}}(t) = 1 - \rho_{\text{LoS}}(t)$.
    
    \subsection{Data Collection Rate in UAV-BD Communications}
    
        For backscatter-based data collection, the UAV transmits an activation RF carrier signal to a BD, and the $k$-th BD backscatters this signal to convey its data back to the UAV. The received signal at the UAV from the $k$-th BD is modeled as
        
        \begin{equation}
            y_k(t) = \sqrt{\xi_{k}} h_{\textrm{f}, k}(t)  h_{\textrm{b}, k}(t) c(t) s_{k}(t) + n(t),
        \end{equation}
        \noindent where $\xi_k$ denotes the backscatter efficiency of the $k$-th BD, $h_{\textrm{f}, k}$ is the forward channel gain (downlink, UAV$\rightarrow$BD) and $h_{\textrm{b}, k}$ is the backscatter channel gain (uplink, BD$\rightarrow$UAV) between the UAV and the $k$-th sensor, respectively.
        With the \textit{monostatic} backscatter configuration with unchanged UAV position during data collection, the forward and backscatter channels can be assumed to be equal, i.e.,
        $h_{\textrm{f}, k}(t) = h_{\textrm{b}, k}(t) \triangleq h_k(t)$.
        $c(t)$ is the UAV's transmitted carrier signal with a power of $ P_{\textrm{c}} = \mathbb{E} \{ \left| c(t)^2  \right| \}$, 
        $s_k(t) \in \{0, 1\}$ represents the baseband OOK signal from $k$-th BD with $\mathbb{E} \{ \left| s_k(t)^2  \right| \} = 1$. 
        The term $n(t)$ represents the additive Gaussian noise with a power of $N_{0}$.
        More specifically, $\xi_{k}$ is related to the polarization mismatch $\chi_{k}$ between antennas of the UAV and the $k$-th BD, the modulation factor $M_{k}$ and on-object penalty $\Theta_{k}$ at the $k$-th BD, which can be expressed by 
        $\xi_{k} = (\chi_{k}^{2} M_{k})/ \Theta_{k}^{2}$~\cite{griffin2009linkbudget}.
        The received carrier signal strength at $k$-th BD is then 
        $P_{\text{r}}^{\text{BD}} = P_{\text{c}} / |h_k(t)|^2$.
        The received backscattered signal strength at the UAV reader is
        $P_{\text{r}}^{\text{TR}} =(\xi_{k} P_{\text{r}}^{\text{BD}})  / |h_k(t)|^2$.
        Hence, the data collection rate $R_k(t)$ for the $k$-th BD is expressed as
        \begin{equation}
            R_k(t) = B \log_2 \left( 1 + \frac{\xi_{k} P_{\textrm{c}} |h_k(t)|^4}{N_0} \right),
        \end{equation}
        where $B$ is the channel bandwidth allocated for each BD, which is considered to be identical among all UAV-BD channels in this work. With $|h_k(t)|^2 = \overline{{L}}(t)^{-1}$, the data collection rate $R_k(t)$ can be also written by
        \begin{equation}
            R_k(t) = B \log_2 \left( 1 + \frac{\xi_{k} P_{\textrm{c}} \overline{L_{k}}(t)^{-2}}{N_0} \right).
        \end{equation}
        
        The time for the data collection of $k$-th BD is $t_{\text{BC}_k} = V_k / R_k(t)$,
        where the $V_k$ is the data volume of this BD. Then, the total time of the communication is denoted as $T_{\text{BC}} = \sum_1^K t_{\text{BC}_k}$.

\subsection{Power Consumption Model}
The total power consumption of the UAV consists of propulsion power for flying $P_{\text{f}}$,  communication power $P_{\text{BC}}$ for exchanging data with BDs, and MA movement power $P_{\text{MA}}$ for the MA control.\par
The propulsion energy $P_{\text{f}}(v)$ required to maintain the UAV in the hovering mode at velocity $v$ can be described as
\begin{equation}
\label{eqn:9}
    \scalebox{0.9}{$
    P_{\text{f}}(v) = P_0 \left( 1 + \frac{3v^2}{v_{\text{tip}}^2} \right) + P_1 \left( \sqrt{1 + \frac{v^4}{4v_0^4}} - \frac{v^2}{2v_0^2} \right)^{\frac{1}{2}} + 
    \frac{\epsilon \Xi \mathcal{S} \mathcal{A} v^3}{2},
    $}
\end{equation}
where $v_{\text{tip}}$ is the rotor tip speed, and $v_0$ represents the average rotor-induced velocity when hovering, $\epsilon$ is the air density, $\mathcal{S}$ and $\mathcal{A}$ represent the solidity and disc area of the rotor in their respective order, $\Xi$ is the fuselage drag ratio.
The parameters $P_0$ and $P_1$ are calculated based on the UAV rotor characteristics and its hovering status
\begin{equation}
    P_0 = \frac{\mu \epsilon \omega^3 \mathcal{R}^3 \mathcal{A}}{8}, \quad P_1 = \frac{(1 + \iota) \mathcal{W}^{\frac{3}{2}}}{\sqrt{2 \epsilon \mathcal{A}}},
\end{equation}
where $\mu$ is the profile drag coefficient, $\epsilon$ is the air density, $\omega$ is the blade angular velocity,  $\mathcal{R}$ is the rotor radius, $\iota$ is the incremental correlation factor to induced power, and $\mathcal{W}$ is weight of the UAV. 
The hovering power of the UAV can be expressed by letting $v=0$ in (\ref{eqn:9}).

The power consumption of the MA is defined as
\begin{equation}
\label{eqn:MApower}
    P_{\text{MA}}(\theta_k, \varphi_k) = P_{\text{base}} + \zeta \cdot \Delta \theta + \kappa \cdot \Delta \varphi,
\end{equation}
\noindent where \( P_{\text{base}} \) represents the minimum power required to initiate and sustain antenna movement, covering static power demands for motor and control systems. $\zeta$ and $\kappa$ are constants for power per unit displacement. $\Delta \theta = |\theta_k - \theta_{\text{prev}}|$, $\Delta \varphi = |\varphi_k - \varphi_{\text{prev}}|$, \( v_{\theta} \) and \( v_{\varphi} \) denote the maximum angular velocities for elevation and azimuth adjustments, respectively. The time $ t_{\text{MA}}(\theta_k, \varphi_k) $ required for the reorientation is
    \begin{equation}
        t_{\text{MA}}(\theta_k, \varphi_k) = \max \left( \frac{\Delta \theta}{v_{\theta}}, \frac{\Delta \varphi}{v_{\varphi}} \right).
    \end{equation}
    \noindent Then, the total time of the MA's orientation is denoted as $T_{\text{MA}} = \sum_1^K t_{\text{MA}}(\theta_k, \varphi_k)$.

\subsection{Problem Formulation}
The UAV's data collection task begins when it departs from its parking station and concludes once data from all BDs has been collected. Our objective is to minimize the task completion time. The problem can be formulated as 

    \begin{equation}
    \begin{aligned}
    \mathbf{P_1:}
    & \min \quad T_{\text{f}} + T_{\text{BC}} + T_{\text{MA}},\\
    \text{s.t.}\ 
    & \mathbf{C1}:  t_{\text{BC}_k}R_k(t) \geq S_k, \\ 
    & \mathbf{C2}:  0 \leq x(t), y(t) \leq L, \\
    & \mathbf{C3}:  0 \leq \theta < \pi/2, \quad 0 \leq \varphi < 2\pi,\\
    & \mathbf{C4}:  P_{\text{r}}^{\text{TR}} \geq P_{\text{sen}}^{\text{TR}}, \\ 
    & \mathbf{C5}:  P_{\text{r}}^{\text{BD}} \geq P_{\text{sen}}^{\text{BD}}, \\
    & \mathbf{C6}:  E_{\text{tot}} \geq E_{\text{f}} + E_{\text{h}} + E_{\text{MA}} + E_{\text{BC}},
    \end{aligned}
    \label{optimization}
    \end{equation}

\noindent where $T_{\text{f}}$ denotes the UAV’s flying time. Since communication and MA movement are performed while hovering, the total hovering time $T_{\text{h}} = T_{\text{BC}} + T_{\text{MA}}$. 
Here, \( E_{\text{tot}} \) denotes the UAV's total energy capacity, with energy components defined as follows: \( E_{\text{f}} = P_{\text{f}}(v) T_{\text{f}} \) for flight, \( E_{\text{h}} = P_{\text{f}}(0) T_{\text{h}} \) for hovering, \( E_{\text{BC}} = P_{\text{BC}} T_{\text{BC}} \) for communication, and \( E_{\text{MA}} = P_{\text{MA}} T_{\text{MA}} \) for MA movement. 
$\mathbf{C1}$ ensures that the data from the $k$-th BD are collected within the allocated time.
$\mathbf{C2}$ restricts the UAV’s position within the boundaries of the target area, while $\mathbf{C3}$ restricts the angles $\theta$ and $\varphi$ for MA orientation.
$\mathbf{C4}$ and $\mathbf{C5}$ ensure that the signal strength received by the UAV reader and BD meet their respective sensitivity thresholds. $\mathbf{C6}$ ensures that the UAV’s total available energy is sufficient to complete the task.

\section{DRL for MA-equipped UAV Data Collection}
\label{sec:DRL}
    This section outlines the SAC-based DRL model developed for the UAV data collection task. It covers the fundamentals of RL, the design of the agent-environment interaction, and details of the SAC model structure and training process. 

    \subsection{Reinforcement Learning Basics}
    RL enables an agent to learn decision-making by interacting with an environment, observing states, taking actions, and receiving rewards to guide future actions. RL problems are often modeled as Markov Decision Processes (MDPs), defined by the tuple $(S, A, P, R, \gamma)$, where $S$ and $A$ are state and action spaces, $P$ is the transition function, $R$ the reward function, and $\gamma$ the discount factor. A transition $\{s, a, r, s'\}$ represents a single step taken by the agent. The agent follows a policy $\pi$ to maximize expected cumulative reward $G(t) = \sum_{i=0}^{T-t-1} \gamma^i r_{t+i+1}$. The goal is to find an optimal policy $\pi^* = \arg\max_\pi \mathbb{E}_\pi \left[ G(t) \right]$ that maximizes this return.

    \subsection{MDP Formulation}
    
    \subsubsection{State Space} 
    The state space provides the RL agent with essential UAV and BD information in a concise form
    \begin{equation} 
        s(t) = \{\mathbf{p_u}(t), \mathbf{o}(t), \mathbf{\Phi}(t), \mathbf{d}(t)\}, 
    \end{equation} 
    \noindent where $\mathbf{p_u}(t) = [x(t), y(t)]$ represents the horizontal coordinates of the UAV. $\mathbf{o}(t) = \{o_0(t), o_1(t), \cdots, o_K(t)\}$ is the set indicating whether the data from each BD has been collected ($o_k(t) = 1$ if collected, $0$ otherwise). $\mathbf{\Phi}(t) = \{\varphi_0(t), \varphi_1(t), \cdots, \varphi_K(t)\}$ gives the azimuth angles of each BD relative to the UAV. $\mathbf{d}(t) = \{d_0(t), d_1(t), \cdots, d_K(t)\}$ represents the horizontal distances of each BD relative to the UAV at this time, where $d_k(t) = \sqrt{(p_x^k - x(t))^2 + (p_y^k - y(t))^2}$.

    \subsubsection{Action Space}
    The action space defines both the UAV’s movement and the MA’s orientation, represented as
    \begin{equation} 
        a(t) = \{a_f(t), a_r(t), \theta_\text{init}(t), \varphi_\text{init}(t)\}, 
    \end{equation}
    \noindent where $a_f(t)$ is the distance factor for the UAV's movement. The actual distance the UAV moves can be calculated as $d_u(t) = a_f(t) \times d_\text{max}$, where $d_\text{max} = \sqrt{L^2 + L^2}$ represents the maximum allowable distance the UAV can move in a single step within the target area. $a_r(t) \in [0, 2\pi]$ represents the direction of the UAV's movement. After executing the action, the UAV's new position $(x_u(t+1), y_u(t+1))$ will be
    \begin{equation}
        \begin{aligned}
            x_u(t+1) &= x_u(t) + d_u(t) \cos(a_r(t)), \\
            y_u(t+1) &= y_u(t) + d_u(t) \sin(a_r(t)).
        \end{aligned}
    \end{equation}
    \noindent To ensure that each UAV movement remains within the target area, the action must satisfy constraint $\mathcal{C}2$ in (15). The action space also includes $\theta_\text{init}(t)$ and $\varphi_\text{init}(t)$, which define the MA's initial elevation and azimuth angles when the UAV arrives at a new data collection position. Upon arriving at a new location, the UAV performs data collection with the MA adjusting to sequentially point toward BDs that meet constraints $\mathcal{C}4$ and $\mathcal{C}5$ in (15). Thus, $a_f(t)$ and $a_r(t)$ determine the UAV’s trajectory, while $\theta_\text{init}(t)$ and $\varphi_\text{init}(t)$ influence the MA adjustment time. Together, these factors impact the overall duration of the data collection task.
    \begin{figure}
    \vspace{-9pt}

    \end{figure}
    \begin{algorithm}[!t]
    \caption{SAC Training}
    \label{alg:sac}
    \begin{algorithmic}[1]
    \State \textbf{Initialize:} Actor \( \pi_{\phi} \), twin Q-networks \( Q_{\theta_1} \), \( Q_{\theta_2} \), target networks \( \theta'_1, \theta'_2 \), replay buffer \( \mathcal{D} \), entropy coefficient \( \alpha \).
    \For{each iteration}
        \State Sample action \( a_t \sim \pi_{\phi}(\cdot | s_t) \), execute \( a_t \), observe \( r_t, s_{t+1} \), store \( (s_t, a_t, r_t, s_{t+1}) \) in \( \mathcal{D} \).
    
        \State \textbf{Sample} mini-batch \( (s, a, r, s') \) from \( \mathcal{D} \).
        \State \textbf{Compute target} Q-value:
        \begin{equation*}
            y = r + \gamma \left( \min_{i=1,2} Q_{\theta'_i}(s', a') - \alpha \log \pi_{\phi}(a' | s') \right)
        \end{equation*}
    
        \State \textbf{Update} Q-networks by minimizing:
        \begin{equation*}
            \mathbb{E}_{(s, a, r, s') \sim \mathcal{D}} \left[ \left( Q_{\theta_i}(s, a) - y \right)^2 \right]
        \end{equation*}
    
        \State \textbf{Update} policy \( \pi_{\phi} \) by maximizing:
        \begin{equation*}
            \mathbb{E}_{s \sim \mathcal{D}, a \sim \pi_{\phi}} \left[ \alpha \log \pi_{\phi}(a | s) - \min_{i=1,2} Q_{\theta_i}(s, a) \right]
        \end{equation*}
    
        \State \textbf{Adjust} \( \alpha \) by minimizing:
        \begin{equation*}
            -\alpha \left( \log \pi_{\phi}(a | s) + \mathcal{H}_{\text{target}} \right)
        \end{equation*}
    
        \State \textbf{Update} target networks: \( \theta'_i \leftarrow \tau \theta_i + (1 - \tau) \theta'_i \).
    \EndFor
    \end{algorithmic}
    \end{algorithm}

    \subsubsection{Reward}
    The reward function provides direct feedback from the environment to guide the UAV in efficient data collection from BDs. It is structured as
    \begin{equation} 
        r(t) = r_b(t) + r_f(t) + p_f(t) + p_{\text{MA}}(t) + p_c(t),
    \end{equation}
    \noindent where $r_b(t) = n_s(t) \times r_{b_{s}}$ represents the reward for collecting data from BDs at the UAV's current position. Here, $n_s(t)$ is the number of BDs satisfying $\mathcal{C}4$ and $\mathcal{C}5$, and $r_{b_{s}}$ is the reward per BD. The condition $r_{b_s} \geq d_\text{max}/v_u$ prevents the UAV from ignoring distant BDs. $r_f(t)$ rewards the UAV for completing data collection from all BDs, encouraging task completion. $p_f(t) = -d_u(t-1) / v_u$ is the penalty for the time spent flying, $p_{\text{MA}}(t)$ represents the penalty for the time spent adjusting the MA, and $p_c(t)$ accounts for the time spent by the UAV in communication with the BDs to collect data; both are calculated for the current $n_s(t)$ BDs. 

    \subsection{Soft Actor-critic Training}
    
    Soft Actor-Critic (SAC) builds on actor-critic (AC) methods, where the agent’s policy (actor) is guided by value estimates (critic) to maximize cumulative rewards. SAC enhances AC by introducing an entropy-regularized objective that balances maximizing rewards with maintaining action randomness, encouraging exploration and avoiding premature convergence \cite{haarnoja2018soft}. This objective, which combines cumulative reward and policy entropy \( \mathcal{H}(\pi(\cdot | s)) \), is expressed as
    \begin{equation}
        J(\pi) = \mathbb{E}_{\pi} \left[ \sum_{t=0}^{T} \gamma^t \left( r(s_t, a_t) + \alpha \mathcal{H}(\pi(\cdot | s_t)) \right) \right],
    \end{equation}
    where \( \alpha \) is the entropy coefficient that adjusts the balance between reward maximization and exploration, with \( \mathcal{H}(\pi(\cdot | s)) = - \mathbb{E}_{a \sim \pi} \left[ \log \pi(a | s) \right] \), encouraging diverse action selection at each state \( s_t \). Thus, SAC enables the agent to effectively explore both the UAV’s trajectory and movable antenna settings during training.
    
    SAC uses two critics, \( Q_{\theta_1} \) and \( Q_{\theta_2} \), which estimate the expected return of a given action in a state. By taking the minimum of the two Q-values, SAC addresses overestimation bias. The target value \( y \) used to update these networks is defined as
    \begin{equation}
    \scalebox{0.78}{$
        y = r(s_t, a_t) + \gamma \, \mathbb{E}_{s_{t+1} \sim p} \left[ \min_{i=1,2} Q_{\theta_i}(s_{t+1}, a') 
         - \alpha \log \pi_{\phi}(a' | s_{t+1}) \right],
         $}
    \end{equation}
    where \( a' \sim \pi_{\phi}(\cdot | s_{t+1}) \) and $p$ represents the environment's state transition probability. Here, \( y \) combines the immediate reward with a conservative estimate of future rewards, adjusted by an entropy term that encourages exploration.
    
    The policy \( \pi_{\phi} \) is updated by maximizing the expected Q-value while balancing it with an entropy term to promote a mix of exploration and exploitation
    \begin{equation}
    \scalebox{0.8}{$
        J(\pi_{\phi}) = \mathbb{E}_{s_t \sim D} \left[ \mathbb{E}_{a_t \sim \pi_{\phi}} \left[ \alpha \log \pi_{\phi}(a_t | s_t) - \min_{i=1,2} Q_{\theta_i}(s_t, a_t) \right] \right],
        $}
    \end{equation}
    where $D$ is the replay buffer storing past experiences for training. To dynamically adjust the level of exploration, the entropy coefficient \( \alpha \) is fine-tuned by minimizing
    \begin{equation}
        J(\alpha) = \mathbb{E}_{a_t \sim \pi_{\phi}} \left[ -\alpha \left( \log \pi_{\phi}(a_t | s_t) + \mathcal{H}_{\text{target}} \right) \right],
    \end{equation}
    where \( \mathcal{H}_{\text{target}} \) regulates exploration, typically set as $-\text{dim}\left(a(t)\right)$ to encourage exploration proportional to action space size. The training process is detailed in Algorithm \ref{alg:sac}.
    
\section{Numerical Results}
\label{sec:Numerical_Results}
    This section presents the numerical results demonstrating the performance of the proposed SAC-based UAV data collection method. Simulation parameters are as follows: The initial positions of the BDs are randomly distributed within the target area, with initial data volumes $S_k$ set randomly between 0.1 and 0.5 Mbits. The UAV’s starting position is set at (0, 0)~m. Other simulation parameters are listed in Table~\ref{tab:parameters}. The data collection efficiency of the proposed directional MA-equipped UAV is compared with that of conventional omni-directional FPA-equipped UAV. The commonly used omni-directional antenna gain of the FPA is set to 5~dBi. Furthermore, we compared the performance of the SAC algorithm with the standard AC algorithm in the data collection tasks involving both types of antennas. \par
        
    \begin{table}[!t]
    \centering
    \caption{Simulation Parameters}
    \resizebox{\columnwidth}{!}{%
    \begin{tabular}{l|l|l}
        \hline
        \textbf{Parameter}      & \textbf{Value}        & \textbf{Description}                     \\ \hline
        $K$                     & 20                      & Default number of BDs                    \\
        $H$                     & 30                      & Default flying height of UAVs [m]        \\
        $L$                     & $200$                   & Default size of target area [m]          \\
        $v_u$                   & 10                      & UAVs' speed [m/s]                         \\
        $f_{\text{c}}$          & 2                       & Carrier frequency [GHz]                  \\
        $P_{\text{sen}}^{\text{TR}}$ & -100           & Reader sensitivity [dBm]  \\
        $P_{\text{sen}}^{{\text{BD}}}$ & -50           
        & BD sensitivity [dBm]  \\
        $B$              & 20                & Operational bandwidth of the UAV [MHz]  \\ 
        $G_{\text{TR}}$     & 10                & Reader antenna gain [dBi]                \\
        $G_{\text{BD}}$     & 0                 & BD antenna gain [dBi]                    \\
        $\chi$           & 0.5               & Polarization mismatch                    \\
        $M$              & 0.5               & Modulation factor                        \\
        $\Theta$         & 0                 & On-object penalty                        \\
        $N_0$            & -100              & Noise power level [dBm]                  \\
       $(\varrho, \beta, \eta_{\text{LoS}},\eta_{\text{NLoS}})$
        & (9.61, 0.16, 1, 20)           & Fading parameters in urban [dB]                \\
        $P_{\text{BC}}$    & 1              & Reader transmission power [W]               \\
        $P_{\text{base}}$    & 2              & Basic MA movement power [W]      \\
        $(\zeta,\kappa)$    & (0.05, 0.03)     &  MA reorientation factors     \\
        $v_{\theta}, v_{\varphi}$    &  $\pi$    &  MA reorientation speed [radians/s]    \\
        $v_{\text{tip}}$   & 80              & Tip speed of the rotor blade [rad/s]                \\
        $v_{0}$            & 5.0463              & Average rotor induced velocity when hovering [m/s] \\
        $\mathcal{R}$   & 0.2              & Rotor radius [m]                                   \\
        $\mathcal{A}$      & 0.1256              & Rotor disc area [m$^2$]                              \\
        $\mathcal{S}$      & 0.1248              & Rotor solidity                                     \\
        $\mathcal{W}$      & 7.84              & UAV weight [N]    \\
        $\mu$              & 0.012              & Profile drag coefficient                          \\
        $\Xi$              & 0.5009              & Fuselage drag ratio                                 \\
        $\omega$           & 400             & Blade angular velocity [rad/s]                        \\
        $\iota$           & 0.05             & Incremental correlation factor to induced power \\
        $\epsilon$         & 1.225              & Air density [kg/m$^3$]             \\
        $r_{b_s}$, $r_f$  & 50, 500   & Reward for data collection \\
        $\gamma$ & 0.99 &  Discount factor \\
        \hline
    \end{tabular}}
    \label{tab:parameters}
    \end{table}


    \begin{figure}[!t]
    \centering
        \includegraphics[width=0.98\columnwidth]{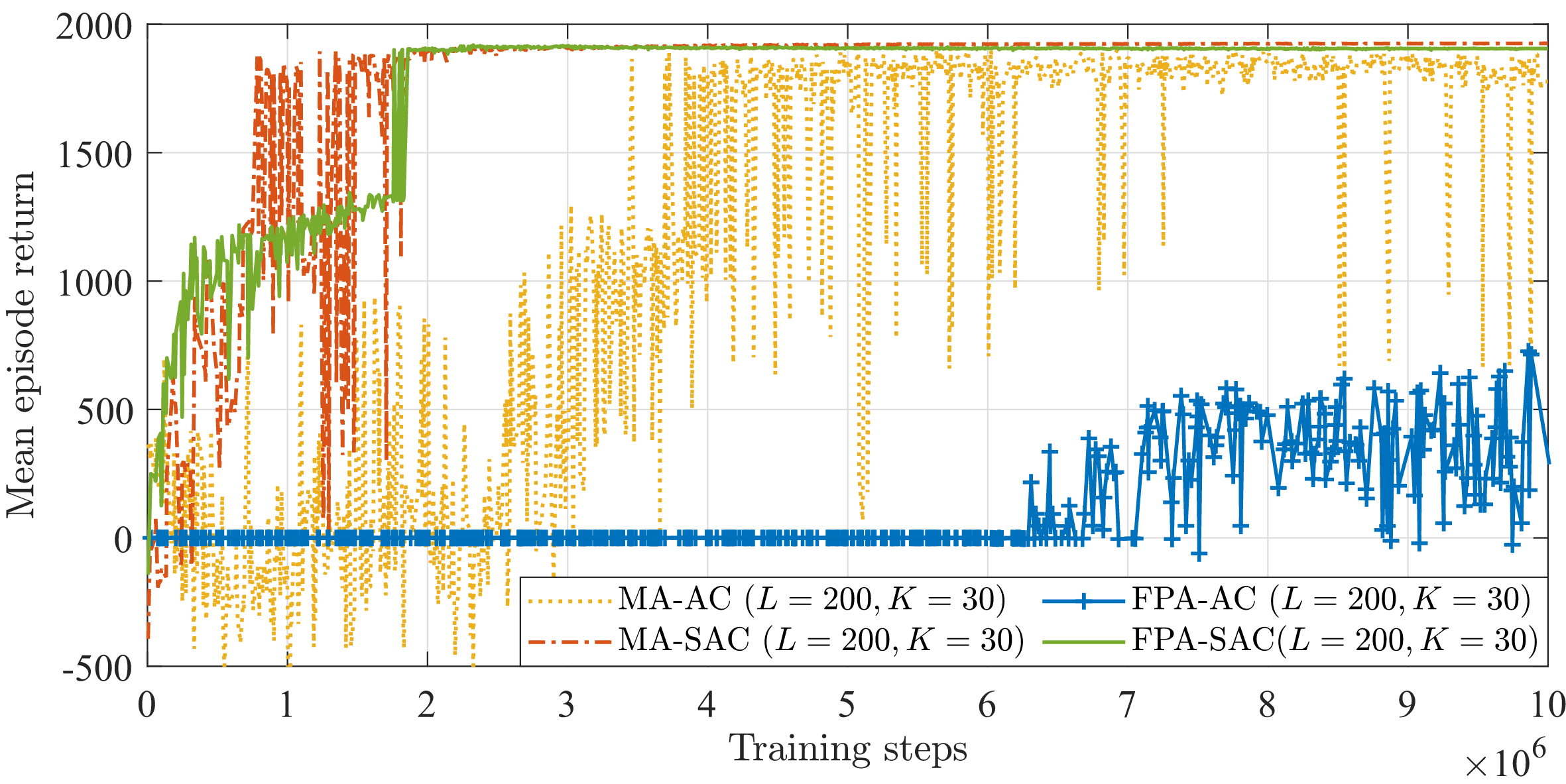}
        \caption{Training convergence of SAC and AC for MA and FPA.} 
        \label{fig:train_curve}
    \end{figure}

    \begin{figure}[!t]
    \centering
        \includegraphics[width=0.98\columnwidth]{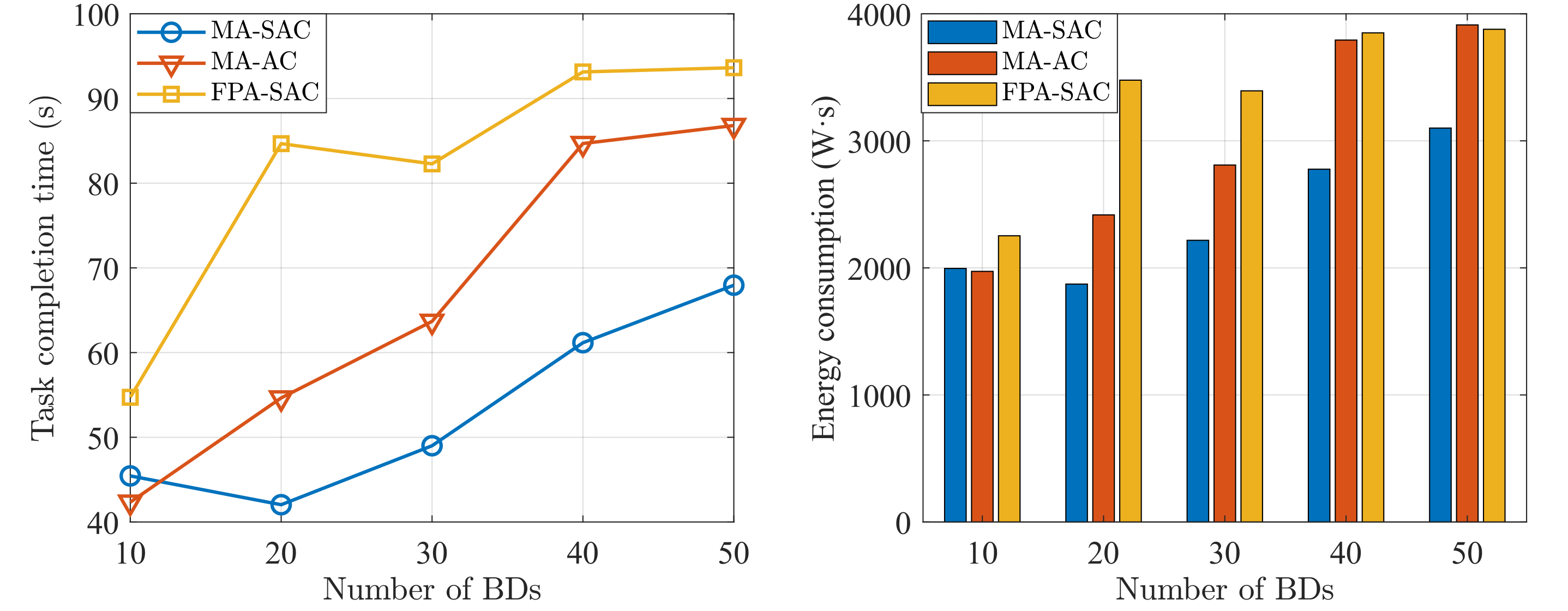}
        \caption{Data collection time and energy consumption versus number of BDs.} 
        \label{fig:bd_time}
         \vspace{-15pt}
    \end{figure}
    Fig.~\ref{fig:train_curve} illustrates a set of training convergence curves for MA-equipped UAV and FPA-equipped UAV, using both SAC and AC algorithms. The x-axis denotes the number of training steps, while the y-axis indicates the mean return per episode. The MA-equipped UAV achieves convergence more efficiently. SAC demonstrates faster, more stable convergence, reaching stability within 3 million steps. In contrast, AC requires around 10 million steps to converge for the MA-equipped UAV and remains less stable, while failing to converge with the FPA-equipped UAV.

\begin{figure}[!t]
    \centering
        \includegraphics[width=0.98\columnwidth]{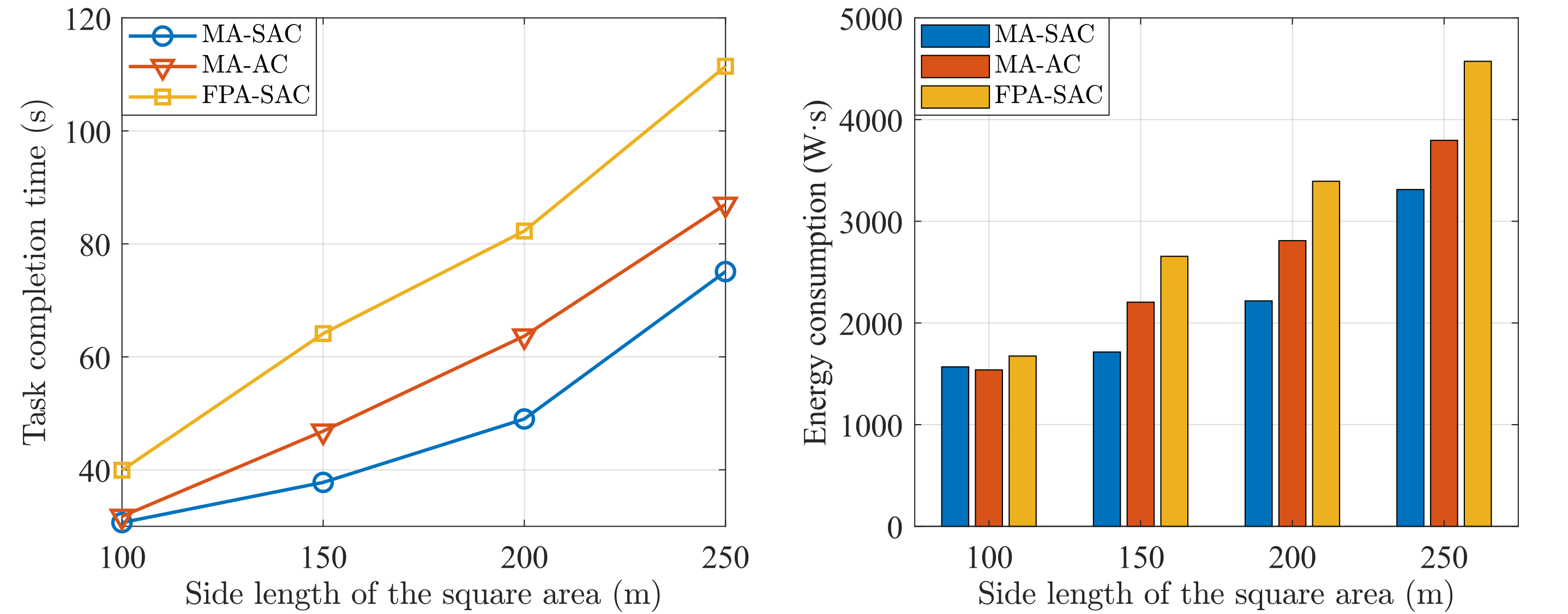}
        \caption{Data collection time and energy consumption versus target area length.} 
        \label{fig:len_time}
    \end{figure}

    Fig.~\ref{fig:bd_time} illustrates the total time and energy consumption for the data collection task as the number of BDs increases, with a fixed area length of $L=200$. The MA-equipped UAV, utilizing the SAC algorithm, shows a notable reduction in both data collection time and overall energy consumption. Furthermore, Fig.~\ref{fig:len_time} presents the total time and energy consumption as the target area size varies, with the number of BDs fixed at $K=20$. Across different area sizes, the MA-equipped UAV demonstrates consistently optimal performance with the SAC algorithm. Due to convergence challenges encountered with the FPA-mounted UAV using the AC algorithm, its results are excluded from these figures.

    \begin{figure}[!t]
    \centering
        \includegraphics[width=0.95\columnwidth]{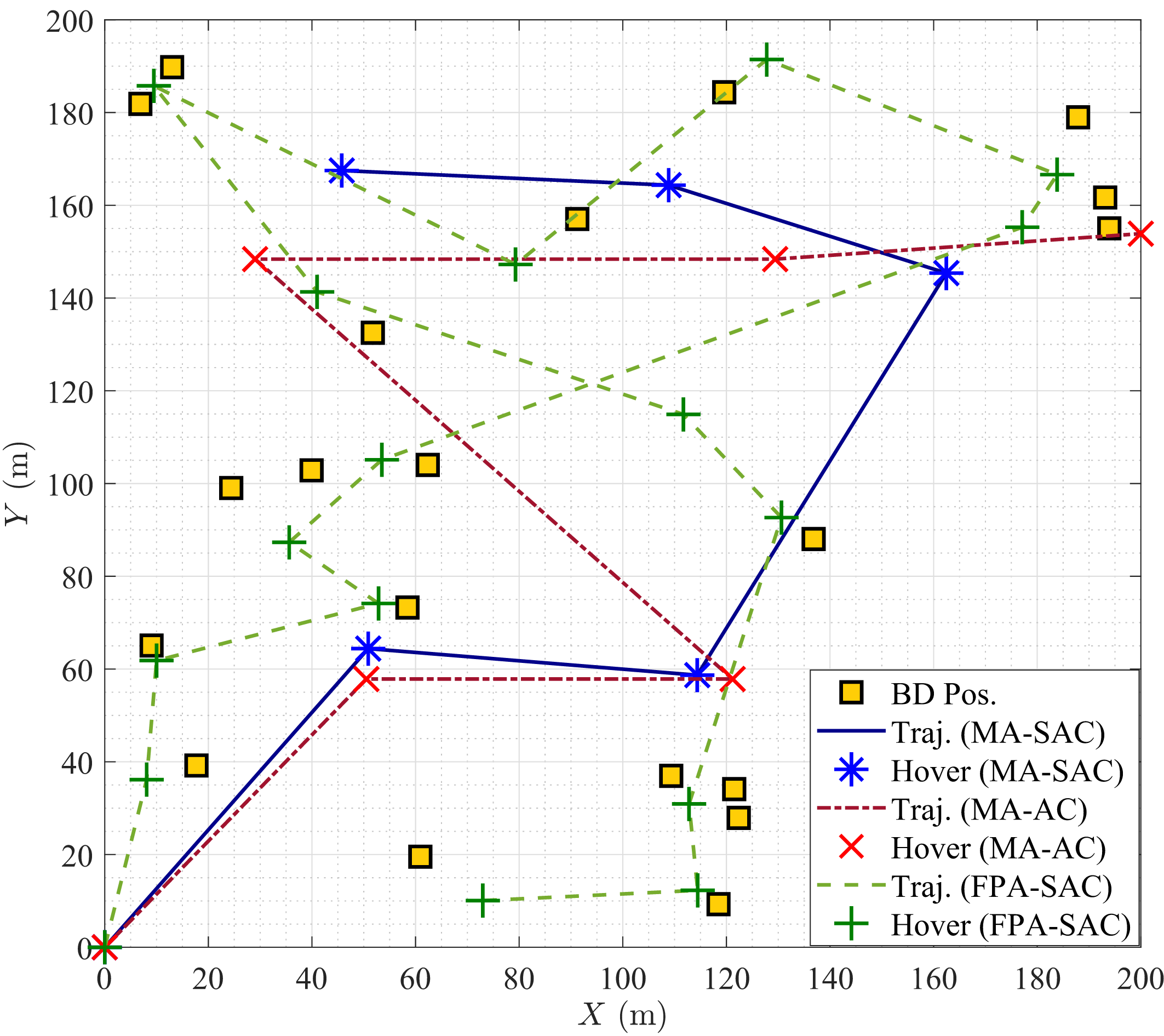}
        \caption{UAV trajectories and hover points for different strategies and antennas.} 
        \label{fig:trajectory}
        \vspace{-15pt}
    \end{figure}
    Fig.~\ref{fig:trajectory} depicts the trajectories of different algorithms and antenna configurations in the data collection task, with parameters set to $L=200$ and $K=20$. The limited directivity and gain of the FPA requires the UAV to fly closer to each BD for establishing communication, leading to near-visit paths to each BD. In contrast, the MA-equipped UAV utilizes antenna rotation to establish communication, reducing the need for extensive UAV movement and resulting in a notably shorter flight distance.
    Moreover, for the MA-equipped UAV, SAC algorithm demonstrates a shorter flying distance and task completion time (365.01~m, 49.00~s) than AC algorithm (447.89~m, 63.69~s).  


\section{Conclusion}
\label{sec:Conclusion}
This paper presents a novel system for enhancing data collection efficiency in backscatter sensor networks by equipping a directional MA on a UAV.
With high directivity and flexibility, the MA enhances communication channel gain by precisely focusing transmission power at each BD.
We minimize the total data collection time by optimizing both the UAV's trajectory and MA's orientation using a SAC-based DRL method. 
Simulation results confirm that the MA-equipped UAV, supported by SAC, outperforms conventional FPA-equipped UAVs and alternative AC algorithms, achieving reduced data collection time and energy consumption.
\bibliographystyle{IEEEtran}
\bibliography{references}

\vspace{12pt}
\end{document}